\documentclass[%
 reprint,
 amsmath,amssymb,
 aip,
 jcp,
floatfix,
amsfonts, amsmath
]{revtex4-1}

\usepackage[version=4]{mhchem}
\usepackage{graphicx}
\usepackage{dcolumn}
\usepackage{bm}

\usepackage[utf8]{inputenc}
\usepackage[T1]{fontenc}
\usepackage{mathptmx}
\usepackage{etoolbox}

\makeatletter
\def\@email#1#2{%
 \endgroup
 \patchcmd{\titleblock@produce}
  {\frontmatter@RRAPformat}
  {\frontmatter@RRAPformat{\produce@RRAP{*#1\href{mailto:#2}{#2}}}\frontmatter@RRAPformat}
  {}{}
}%
\makeatother

\newcommand{\eff}{{\Lambda}}
\newcommand{\proj}[1]{\Theta_{#1}}
\newcommand{\ecore}{E_{\text{core}}}

\begin{document}


\title{Perturbation-Adapted Perturbation Theory}

\author{Peter J. Knowles}
 \email{KnowlesPJ@Cardiff.ac.uk}
\affiliation{School of Chemistry, Cardiff University, Main Building, Park Place, Cardiff CF10 3AT, United Kingdom}%

\date{\today}

\begin{abstract}
A new general approach is introduced for defining an optimum zero-order Hamiltonian for Rayleigh-Schr\"odinger perturbation theory. Instead of taking the operator directly from a model problem, it is constructed to be a best fit to the exact Hamiltonian within any desired functional form. When applied to many-body perturbation theory for electrons, strongly improved convergence is observed in cases where the conventional Fock Hamiltonian leads to divergence or slow convergence.
\end{abstract}

\maketitle


Rayleigh-Schr\"odinger perturbation theory (RSPT)\cite{Schrodinger1926} is a ubiquitous tool for the simplification and analysis of many problems in quantum mechanics.
We here consider its application to the ground-state eigenvalue of the time-independent Schr\"odinger equation.
In its usual form, the first step is to identify a model problem that is in some sense close to the target problem, but for which exact solutions to Schr\"odinger's equation can be obtained.
The energy and wavefunction for the target problem are then obtained recursively as a power series in the perturbing Hamiltonian, defined as the difference between the target- and model-problem Hamiltonians.
The process is usually formulated under the assumption that the complete set of eigenfunctions of the model problem is available. Success is defined by realizing sufficient accuracy from the series for the energy truncated at low order.

It has long been recognized that the meaningfulness of low-order truncation of perturbation series is dependent at least to some extent on the existence and rapidity of convergence of the series.
Although even when a perturbation series is formally divergent, the first terms of an asymptotic expansion can still be useful, it is generally the case that divergence or slow convergence mean that the low-order energies may not be accurate enough to be useful approximations to the true Hamiltonian eigenvalue. In some instances, convergence can be improved by resummation\cite{Wilson1976,Silver1977,Sadlej1981,Goodson2000,Goodson2002,Roth2010,Goodson2012,mihalka2017,Mihalka2017a,Mihalka2019} or scaling\cite{Schmidt1993b,Sadlej1981} techniques, but in general these are not straightforward to apply in a way that preserves extensivity of the energy in many-body theory.

Although the model problem may have a physical reality that aids interpretation and which defines the model Hamiltonian operator $\hat H_0$ naturally, there is no intrinsic requirement for either of these for RSPT to be effective.
What, however, is essential is that one has an operator whose ground state is a zero-order wavefunction that is a good approximation to the ground state of the target problem; there is complete flexibility about the remainder of its spectrum.
We explore here this flexibility, and propose a general 
procedure for defining an optimum $\hat H_0$ given a particular zero-order wavefunction $|0\rangle$ and exact Hamiltonian $\hat H$, together with any constraints on the form of $\hat H_0$ necessary for sufficiently simple computation.

RSPT starts by choosing an initial zero-order hamiltonian
$\hat H_{0}$, which then defines the initial perturbation operator $\hat H_{1}=\hat H-\hat H_{0}$. Schr\"odinger's equation is then partitioned for all values of the perturbation strength $\lambda$,
\begin{align}
    \sum_{n=0}\lambda^n\Big(
    \left(\hat H_{0}-E_{0}\right)|n\rangle
    +\left(\hat H-\hat H_{0}-E_{1}\right)|n-1\rangle
    \nonumber\\
    -\sum_{m=0}^{n-2} E_{n-m}|m\rangle
    \Big)=0
    \label{eq:tise}
\end{align}
which can be solved order by order in the usual way to obtain the energy and wavefunction.

We now seek an improved zero-order Hamiltonian, $\hat \eff$, by maximizing the degree to which it resembles the full Hamiltonian, as measured by its action on the first-order wavefunction projected onto a suitable space.
The rationale for this approach is that it is known that sometimes the convergence of RSPT can be spoiled if the eigenvalues of $\hat H_0$ are markedly different from those of $\hat H$; it can then happen that at some point inside the unit circle in the complex plane, $\hat H_0+\lambda(\hat H-\hat H_0)$, acquires a degenerate ground state, and then at $\lambda=1$ the perturbation series is divergent\cite{Christiansen1996b,Olsen2000b,Goodson2004,Goodson2006,Sergeev2006,Olsen2019,Marie2021a}. Slow convergence can usually also be associated with near degeneracy somewhere in the unit circle.
The similarity conditions can be expressed as
\begin{align}
    \langle 0| \hat\Theta_p^\dagger (\hat \eff-\hat H) | 1\rangle &=0
    \label{eq:proj}
\end{align}
where $\{\hat\Theta_p\}$ is an appropriately-chosen set of operators.
It is anticipated that $\hat \eff$ will be defined through a number of adjustable parameters to be determined from the solution of (\ref{eq:proj}).
We denote the power-series expansion of energy and wavefunction based on this partitioning Perturbation-Adapted Perturbation Theory (PAPT), since the zero-order hamiltonian is no longer universal, but depends on the nature and strength of the perturbation.

An obvious choice for the improved zero-order hamiltonian is $\hat \eff=\hat H$.  For most problems, this will not offer a practical advantage to simply solving Schr\"odinger's equation exactly; the linear eigenvalue problem is replaced by a sequence of inhomogeneous linear equations of the same dimension. But we will explore the properties of the approach by applying it to the perturbed 1-dimensional harmonic oscillator. We also note that this choice of the exact Hamiltonian is related to Epstein-Nesbet perturbation theory\cite{Epstein1926,Nesbet1955}, which incorporates the further approximation of ignoring the off-diagonal elements of $\hat \eff$ for computational simplicity.

More usually, we will construct $\hat \eff$ to have the same functional form as $\hat H_{0}$, with linear adjustable parameters. We consider below the case of electronic structure theory with the Fock $\hat H_{0}$; in that case, to retain computational efficiency, $\hat \eff$ should also be a one-body operator whose spectrum mimics that of the two-body $\hat H$.

\section{Application to the perturbed harmonic oscillator}
We consider a zero-order hamiltonian for the harmonic oscillator with unit mass and force constant, and taking $\hbar=1$:
\begin{align}
    \hat H_{0} &= -\tfrac12 \frac{d^2}{dx^2} + \tfrac12 x^2
\end{align}
which has eigenvalues $(n+\tfrac12), n=0,1,2,\dots$ and known eigenfunctions.
We then add a damped quartic perturbation,
\begin{align}
    \hat H &= \hat H_0+\lambda\, x^4 e^{-x^2/8}
\end{align}
It is well known that without the gaussian damping factor, perturbation theory diverges for all $\lambda\ne 0$,
associated with the fact that for any finite negative $\lambda$ there is a negative-energy continuum of eigenstates.

In order to develop PAPT for the ground state of this system, we require that the harmonic-oscillator ground state is an eigenfunction of $\hat \eff$, but that otherwise $\hat \eff$ is as close as possible to the actual Hamiltonian $\hat H_{0}+\lambda(\hat H-\hat H_{0})$. This can be achieved by simple projection,
\begin{align}
    \hat \eff &= 
   \hat P
    \hat H_{0}
    \hat P
    +
    (1-\hat P)(\hat H_{0} +\lambda (\hat H-\hat H_{0}))(1-\hat P)
\end{align}
where $
   \hat P = |0\rangle \langle 0|
$.
\begin{table}[ht]
    \centering
    \begin{tabular}{c|d|d}
    \hline
    \hline
    Order & \multicolumn1{c|}{RSPT} & \multicolumn1c{PAPT} \\
    \hline
    \hline
    1 &7.53\times 10^{-2} & 7.53\times 10^{-2} \\
    2 & -2.30\times 10^{-3} & -2.19\times 10^{-5} \\
    3 & 8.91\times 10^{-4} & -2.19\times 10^{-5} \\
    4 & -4.63\times 10^{-4} & 1.303 \times 10^{-7} \\
    5 & 1.74\times 10^{-4} & 1.303 \times 10^{-7} \\
    6 &-1.43\times 10^{-4} & 1.303 \times 10^{-7} \\
    7 &2.01\times 10^{-5} & -9.73 \times 10^{-10} \\
    \hline
    \end{tabular}
    \caption{Deviation from exact of truncated perturbation series for the ground-state energy of the damped-quartic-perturbed harmonic oscillator with $\lambda=0.1$ at different truncation orders.}
    \label{tab:quartic}
\end{table}
Table~\ref{tab:quartic} shows the convergence of RSPT and PAPT as measured by the remaining difference from the exact ground-state eigenvalue after truncating the perturbation series at different orders.  The calculations have been carried out using a basis consisting of the first 30 harmonic-oscillator eigenfunctions. RSPT is very slowly convergent, whereas PAPT converges rapidly. It is also possible to obtain similarly rapid convergence of PAPT for the undamped perturbation $\lambda\,x^4$, for which RSPT diverges for all $\lambda$.
We note that in this application of PAPT using the full hamiltonian, the third-order energy (and systematically other higher-order contributions) is zero because the representation of $\hat H-\eff$ in the first-order interacting space is zero.
\section{Many-body perturbation theory for electrons}
Conventional RSPT for the electron correlation problem proceeds using the M\o ller-Plesset zero-order Hamiltonian\cite{Mller1934,Binkley1975}
\begin{align}
    \hat H_{0} &=
 \hat f = f_{ab}\,a^\dagger b - f_{ij}\,j i^\dagger   
 +f_{ii} +\ecore
\end{align}
where $i, i^\dagger, a, a^\dagger$ are the usual annihilation and creation operators for, respectively, occupied and virtual orbitals, $\hat f$ is the Fock operator of the reference Hartree-Fock calculation, $\ecore$ is the nuclear-nuclear Coulomb energy, and the Einstein summation convention is used.

We now seek
\begin{align}
 \hat \eff &= \eff_{ab}\,a^\dagger b - \eff_{ij}\,j i^\dagger   
 +\eff_{ii} +\ecore
\end{align}
such that
\begin{align}
    \hat\eff|1\rangle &\approx \hat H|1\rangle
    \label{eq:effective}
\end{align}
where  $|1\rangle$ is the first-order RSPT wavefunction,
\begin{align}
|1\rangle&=
\tfrac14 c^{ij}_{ab}
    \,
    a^\dagger b j i
    |0\rangle
\end{align}
To find the parameters $\{\eff_{ab}\},\{\eff_{ij}\}$, we project (\ref{eq:effective}) onto a suitable space. The vectors 
\begin{align}
|\proj{ij}\rangle &=\tfrac14 c^{ik}_{ab} \,a^\dagger b^\dagger k j |0\rangle
\\
|\proj{ab}\rangle &=\tfrac14 c^{ij}_{ac} \,b^\dagger c^\dagger j i |0\rangle
\end{align}
are equal in number to the unknowns, and sample the first-order interacting space.
The idea is that the reference function plus first-order interacting space contain the dominant part of the wavefunction, and that many higher excited states will arise as unlinked clusters from this space, and so if the similarity condition is obeyed in the first-order space (or even a subspace), it is likely to be reasonable in the full Hilbert space too.
The projection vectors contain one linear dependency, $|\proj{ii}\rangle = |\proj{aa}\rangle=|10\rangle$,
and so 
 projection of (\ref{eq:effective}) onto
$\{|\proj{ij}\rangle\}\oplus\{|\proj{ab}\rangle\}$ results in
a linear equation system with one redundant equation, which matches a redundant parameter in  the parameters:
the perturbation equations
depend only on differences of eigenvalues of $\eff$, and are therefore unaffected by a constant shift in all of its diagonal elements.

We further require that $\hat\eff$ is a physically reasonable, and therefore Hermitian, operator. This can be achieved by enforcing $\hat\eff=\hat\eff^\dagger$ and projecting against the space
$\{|\proj{ij}+\proj{ji}\rangle, i\ge j\}\oplus\{|\proj{ab}+\proj{ba}\rangle,a\ge b\}$.

\begin{figure}
    \centering
    \includegraphics[width=\linewidth]{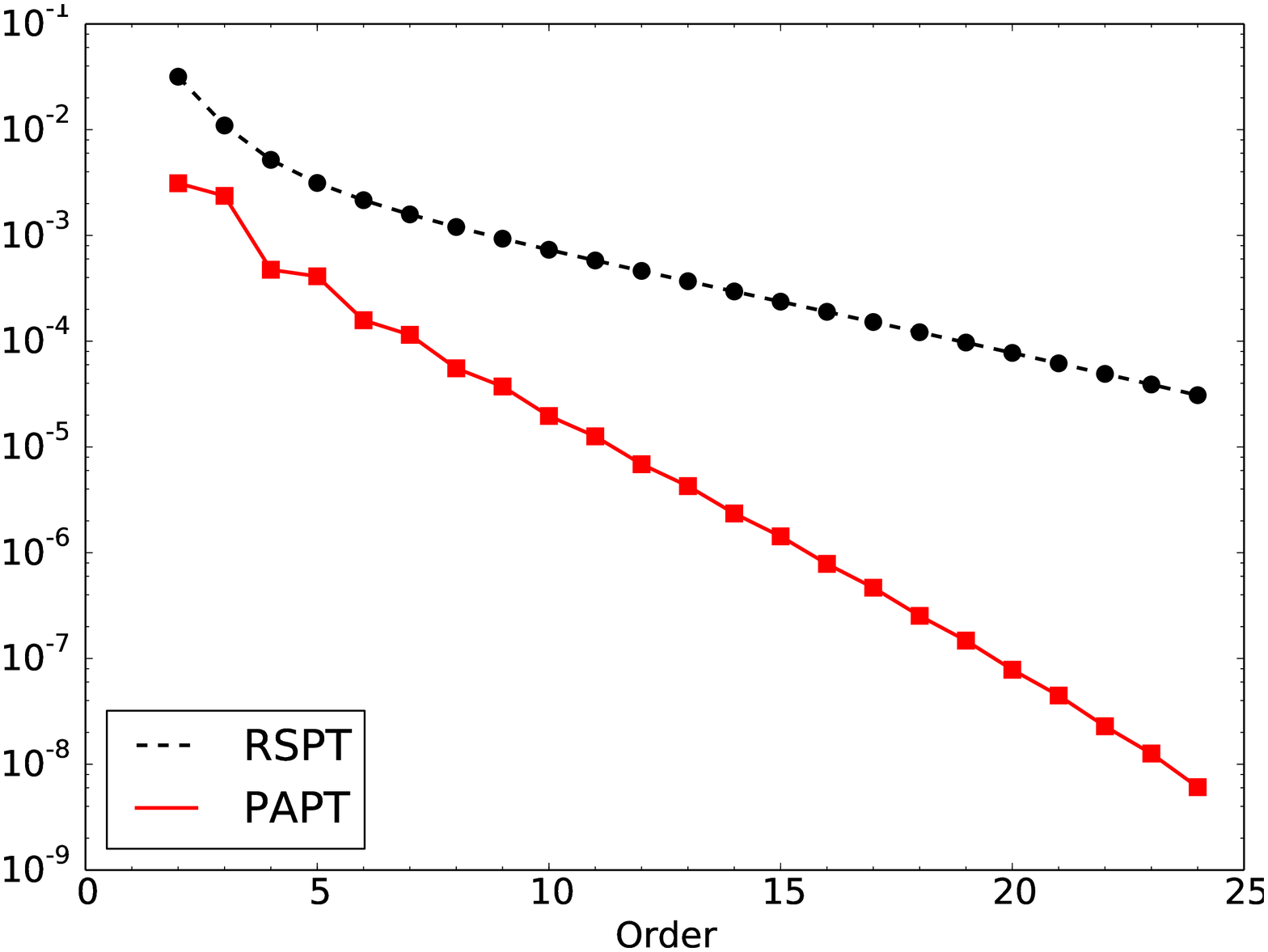}
    \caption{Deviation from full configuration interaction of truncated perturbation series for the ground-state energy / Hartree of $^{1}{A}_1$ \ce{CH2}  at different truncation orders.}
    \label{fig:ch2}
\end{figure}
Figure~\ref{fig:ch2} compares the performance of PAPT and conventional M\o ller-Plesset (MP) RSPT for the ground singlet state of the \ce{CH2} molecule (bond length $1.102\,\text{\AA}$, angle $104.6^\circ$). This is an example in which a low-lying electronically-excited state causes convergence of RSPT to be slow. Using the cc-pVDZ basis set\cite{DunningJr1989a}, and excluding excitations from the lowest (C 1s) orbital, full configuration interaction calculations are possible, and define the exact problem. With the full Slater-determinant basis, perturbation theory is then propagated to any desired order\cite{Knowles1985,Handy1985}. The difference between the perturbation series for the energy, truncated at successive orders, and exact is plotted as measure of degree of convergence. The use of a logarithmic scale reflects the fact that both converge monotonically and exponentially with order, but it is seen that the convergence of PAPT is significantly better than that of RSPT, with errors that are more than an order of magnitude smaller at all orders, and a much faster asymptotic rate of convergence.

\begin{figure}
    \centering
    \includegraphics[width=\linewidth]{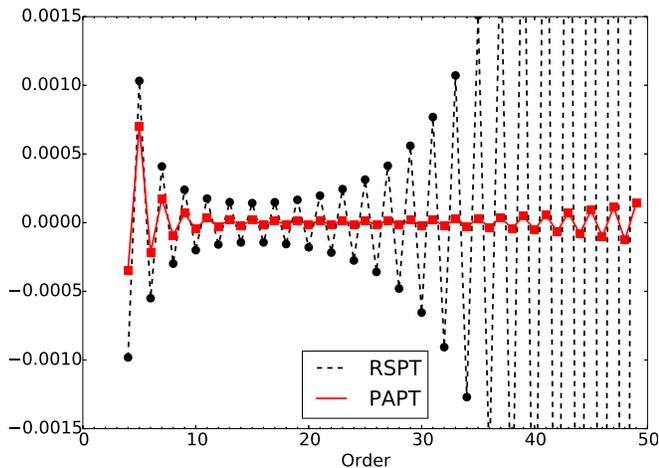}
    \caption{Deviation from full configuration interaction of truncated perturbation series for the ground-state energy / Hartree of \ce{Ne}  at different truncation orders.}
    \label{fig:ne}
\end{figure}
Figure~\ref{fig:ne} shows the same measures for the Ne atom in the aug-cc-pVDZ\cite{DunningJr1989a,Kendall1992} basis, and with excitations from the $1s$ orbital omitted. This has already been established\cite{Christiansen1996b} as an example of divergence arising from ``back-door intruders'', ie highly-excited states which for certain perturbation strengths within the unit circle in the complex plane, and with negative real part, become degenerate with the ground state\cite{Christiansen1996b,Olsen2000b,Goodson2004,Goodson2006,Sergeev2006,Olsen2019}.
Not shown on the figure are the third-order errors, which are similar for PAPT (0.00417) and RSPT (0.00470). The RSPT series is oscillatory, and the energies begin to diverge from around 16th order. The PAPT series is also oscillatory, and formally divergent, but divergence begins at higher order, and when the errors are already much smaller.


These examples exhibit a third-order energy that is small but non-zero, and generally smaller contributions from odd orders than from even. This is a consequence of $\hat \eff$ reproducing $\hat H$ in the first-order space generated by $\hat H-\hat H_0$, which is near, but not identical, to that arising from $\hat H-\hat\eff$.
This illustrates a connection between PAPT and Feenberg scaling\cite{Schmidt1993b}, in which $\hat H_0$ is scaled to make the third order energy zero. The Feenberg approach defines a single global scaling parameter, which unfortunately leads to a lack of extensivity, whereas in PAPT, which is defined entirely through linked tensor contractions, extensivity is preserved.
One could envisage proceeding further by making an improved zero-order Hamiltonian by using the PAPT first-order wavefunction instead of $|1\rangle$ in equation (\ref{eq:effective}), and even iterating until self-consistent, giving a zero third-order energy, and reduced error at second order.  The computational cost for each determination of an improved zero-order Hamiltonian is essentially the same as a third-order calculation; therefore PAPT3 is roughly twice the cost of MP3, and it is probably not worthwhile to iterate further. At higher order, the overhead for PAPT is essentially negligible.

\section{Conclusion}
It is possible to significantly improve the convergence characteristics of Rayleigh-Schr\"odinger perturation series by adopting a zero-order Hamiltonian that is parameterized to mimic the full Hamiltonian in the first-order interacting space.
The method is aspirationally similar to resummation schemes, as well as Similarity Renormalization Group and canonical transformation approaches\cite{Glazek1994,White2002,Evangelista2014c}, in which singularities arising from actual or near degeneracy are removed by appropriate transformations, but is defined entirely through linear equations.
In general, the approach may not offer a significant practical advantage if the resulting zero-order Hamiltonian does not have a simple form.
In the special case of electronic structure theory, we have shown that it is possible to find a one-electron operator that represents the full two-electron Hamiltonian, at the cost of an additional third-order perturbation treatment. In the examples considered, the convergence of the perturbation series is substantially improved. 

\bibliography{PJK-RSPT-convergence}

\end{document}